\documentstyle[amssymb,aps,prl,epsfig]{revtex}
\begin{document}
\draft
\twocolumn[\hsize\textwidth\columnwidth\hsize\csname %
@twocolumnfalse\endcsname

\title{Spectral functions, Fermi surface and pseudo gap in the  
$t$-$J$ model}   
\author{P. Prelov\v sek and A. Ram\v sak}
\address{Faculty of Mathematics and
Physics, University of Ljubljana, 1111 Ljubljana, Slovenia }
\address{J. Stefan Institute, University of Ljubljana, 1111 Ljubljana,
Slovenia }
\date{September 4, 2001}
\maketitle 
\begin{abstract}
\widetext

Spectral functions within the generalized $t$-$J$ model as relevant to
cuprates are analyzed using the method of equations of motion for
projected fermion operators. In the evaluation of the self energy the
decoupling of spin and single-particle fluctuations is performed. It
is shown that in an undoped antiferromagnet (AFM) the method
reproduces the selfconsistent Born approximation. For finite doping
with short range AFM order the approximation evolves into a paramagnon
contribution which retains large incoherent contribution in the hole
part of the spectral function as well as the hole-pocket-like Fermi
surface at low doping. On the other hand, the contribution of
(longitudinal) spin fluctuations, with the coupling mostly determined
predominantly by $J$ and next-neighbor hopping $t'$, is essential for
the emergence of the pseudogap. The latter shows at low doping in the
effective truncation of the large Fermi surface, reduced electron
density of states and at the same time quasiparticle density of states
at the Fermi level.

\end{abstract}
\pacs{PACS numbers: 71.27.+a, 72.15.-v, 71.10.Fd} ]
\narrowtext

\section{Introduction}

One of the central issues in the experimental and theoretical
investigations of superconducting cuprates is the understanding of
low-energy electronic excitations in these compounds \cite{imad},
where the clue to their high-temperature superconductivity lies
possibly already in their anomalous normal-state properties. In recent
years, in particular, the remarkable progress in the angle-resolved
photoemission spectroscopy (ARPES) experiments revealed quite
universal development of electron spectral properties as a function of
doping. In most investigated Bi$_2$Sr$_2$CaCu$_2$O$_{2+\delta}$
(BSCCO) ARPES shows quite well defined large Fermi surface in the
overdoped and optimally doped samples at $T>T_c$, whereby the
low-energy behavior with increasing doping in the overdoped regime
qualitatively approaches (but does not in fact reach) that of the
normal Fermi-liquid with underdamped quasiparticle (QP)
excitations. On the other hand, in the underdoped materials the QP
dispersing through the Fermi surface (FS) are resolved by ARPES in
BSCCO only in parts of the large FS, in particular along the nodal
$(0,0)$-$(\pi,\pi)$ direction \cite{mars}, indicating that the rest of
the large FS is truncated \cite{norm}, i.e. either fully or
effectively gaped.  At the same time near the $(\pi,0)$ momentum
ARPES reveals a hump at $\sim 100$~meV \cite{mars}, which indicates on
the existence of the pseudogap scale, which is consistent with the
characteristic temperature pseudogap scale $T^*>T_c$, which appears
also as a crossover in several other quantities: uniform
susceptibility $\chi(T)$, resistivity $\rho(T)$, the specific heat
$C_V(T)$ and the Hall constant $R_H(T)$ \cite{imad}. Although the
latter anomalies in thermodynamic and transport quantities are quite
similar (or even better confirmed and more pronounced) also in
La$_{2-x}$Sr$_x$Cu$_2$O$_4$ (LSCO), spectral properties of the latter
\cite{fuji} are qualitatively different from BSCCO, presumably due to
the crucial role of stripe structures in the LSCO in the regime of
intermediate doping.

Since electron spectral functions are the central quantity to
characterize and understand anomalous Fermi liquid in cuprates, they
have been the subject of numerous theoretical studies. There appears
to be at least a theoretical consensus on the spectral functions in an
undoped reference antiferromagnet (AFM), describing a single hole
inserted in an AFM behaving in two-dimensional (2D) planar system as a
QP with strongly renormalized mass and large incoherent component. The
spectral function is well captured within the selfconsistent Born
approximation (SCBA) \cite{kane} for the simplest relevant $t$-$J$
model, whereby for an agreement with experiments on undoped cuprates,
i.e., Sr$_2$CuO$_2$Cl$_2$ \cite{well} and Ca$_2$CuO$_2$Cl$_2$ \cite{ronn}
longer range hopping terms have been invoked \cite{dago,tohy}.

For larger (finite) doping in the lack of reliable analytical
techniques numerical approaches have been used extensively. Starting
with the simplest models for correlated electrons in cuprates, i.e.,
the Hubbard model and the $t$-$J$ models on planar lattices, numerical
studies employing mainly the exact-diagonalization and the quantum
Monte Carlo methods were able to confirm some gross features
consistent with experiments.  In particular this includes: a) the
existence of the large FS already in moderately doped AFM
\cite{step,dago}, b) the overdamped character of QP at the
intermediate doping \cite{jpspec,jprev}, consistent with the marginal
Fermi-liquid concept \cite{varm}, c) the pseudogap features at lower
doping in spectral functions \cite{preu} and in the density of states
(DOS) \cite{jprev}, and d) quite visible contribution of longer range
hopping \cite{tohy}. Whereas numerical studies confirm the relevance
of simplest models and results for cuprates, still they are prevented
to approach effectively the low-energy regime and results need a
proper analytical (phenomenological) interpretation.

Analytical approximations to spectral properties of relevant models
for 2D systems at finite doping have proven to be very delicate due to
strong interaction between electron excitations, spin degrees and
pairing fluctuations. For the one-band Hubbard model spectral
functions have been evaluated within the random-phase approximation
for AFM fluctuations \cite{kamp} and within the self-consistent
conserving (FLEX) theory \cite{bick}, both restricted to modest
$U/t$. Starting with the $t$-$J$ model strong correlations are
explicitly taken into account in slave boson theories \cite{wang},
where it is however difficult to incorporate AFM spin
fluctuations. The latter play the essential role in phenomenological
theory of the spin-fermion model leading to the nearly AFM Fermi
liquid \cite{mont}.  Recently aspects of the pseudogap features in the
underdoped regime have been found in this model evaluating the self
energy involving a strong coupling to AFM spin fluctuations
\cite{chub,schm}.

Concerning the origin and the explanation of the pseudogap scale is
seems plausible that at low doping it is related to the exchange $J$
since $T^*\sim J$ in low-doping materials, whereas it merges $T^* \sim
T_c$ in optimally doped samples. This indicates on the importance of
AFM spin correlations for the emergence of the (large) pseudogap as
found also in the numerical studies \cite{preu,jpspec,jprev} and in
phenomenological model studies \cite{chub,schm}. The renormalization
group studies of the Hubbard model \cite{zanc} (with moderate $U/t$)
also reveal the instability of the normal Fermi liquid close to the
half-filled band (insulator) and a possible truncation of the Fermi
surface, but here either the spin or pairing fluctuations can take the
dominant role depending on FS nesting conditions whereby the
longer-range hopping terms can be quite important.
 
One of present authors introduced the equations-of-motion (EQM) method
\cite{prel} for the evaluation of the spectral functions within the
$t$-$J$ model. It has been shown that EQM for projected fermionic
operators implicitly reveal an effective spin-fermion coupling. Using
the simplest decoupling it was possible to relate the overdamped
marginal-type character of QP to the marginal dynamics of spins
\cite{prel} but also to treat the superconducting fluctuations and
transition \cite{plak}. EQM method has been also applied for the
Hubbard model \cite{onod}. The analysis of spectral functions within
the $t$-$J$ model has been recently improved \cite{prel1} by more
appropriate treatment of the self energy by dealing separately with:
a) the strong coupling to transverse short-range AFM spin fluctuations
- paramagnons and b) the moderate coupling to slower longitudinal
fluctuations of the AFM order parameter. The theory and results were
partially presented already in Ref.\cite{prel1}. The aim of this paper
is to present the theory and in particular its consequences and
results in more detail. The main advantage of this theory is that it
allows for the study of the evolution of spectral function as a
function of doping: from an undoped system up to moderate doping. The
emphasis is on the results at $T=0$, in particular on: a) the
development of the FS from a hole-pocket like into a large one, b) the
emergence of the pseudogap in the spectral function and related
effective truncation of the FS most pronounced near $(\pi,0)$ in the
Brillouin zone, c) anomalous properties of QP at the FS remaining well
defined even in the pseudogap regime, d) depleted DOS and moreover the
quasiparticle DOS (related to specific heat coefficient) with doping.

The paper is organized as follows: In Sec.~II EQM method for the
spectral function within the $t$-$J$ model is summarized. Sec.~III is
devoted to the evaluation of the self energy within the decoupling
approximation separately treating the paramagnon contribution
$\Sigma_{\rm pm}$ and the contribution of longitudinal spin fluctuations
$\Sigma_{\rm lf}$. In Sec.~IV results of the simplified analysis of the
pseudogap features are presented taking into account an effective
renormalized band and explicitly $\Sigma_{\rm lf}$. Sec.~V presents results
of the full selfconsistent solution for spectral function as a function
of doping.

\section{Equations of motion}

In order to take explicitly into account strong correlations we study
the $t$-$J$ model,
\begin{equation}
H=-\sum_{i,j,s}t_{ij} \tilde{c}^\dagger_{js}\tilde{c}_{is}
+J\sum_{\langle ij\rangle}({\bf S}_i\cdot {\bf S}_j-\frac{1}{4}
n_in_j) , \label{eq1}
\end{equation}
where fermionic operators are projected ones not allowing for the
double occupancy of sites, i.e., 
\begin{equation}
\tilde{c}^\dagger_{is}= (1-n_{i,-s}) c^\dagger_{is}. \label{eq2}
\end{equation}
Since longer range hopping appears to be important for the proper
description of spectral function in cuprates, both for the shape of
the FS at optimum doping materials \cite{tohy} as well as for the
explanation of ARPES of undoped insulators \cite{well,ronn,tohy}, we
consider besides $t_{ij}=t$ for the n.n. hopping also $t_{ij}=t'$ for
the n.n.n. hopping on a square lattice.

Our goal is to evaluate the electron Green's function (propagator)
directly for projected fermionic operators,
\begin{eqnarray}
&&G({\bf k},\omega)= \langle\!\langle \tilde c_{{\bf k} s}; \tilde
c^{\dagger}_{{\bf k} s} \rangle\!\rangle _{\omega} =\nonumber \\
&&= -i \int_0^{\infty}
{\rm e}^{i(\omega+\mu ) t}\langle \{ \tilde c_{{\bf k} s}(t) ,
\tilde c^{\dagger}_{{\bf k} s} \}_+  \rangle dt, 
\label{eq3}
\end{eqnarray}
which is equivalent to the usual propagator within the allowed basis
states of the model, Eq.~(\ref{eq1}). In the EQM method \cite{zuba} one
uses relations for general correlation functions
\begin{eqnarray}
\omega \langle \!\langle A;B \rangle \!\rangle_\omega &=& \langle
 \{A,B\}_+\rangle + \langle \!\langle [A,H]; B \rangle
\!\rangle_\omega = 
\nonumber \\
 &=& \langle \{A,B\}_+\rangle - \langle \!\langle A;[B,H]
 \rangle \!\rangle_\omega.
\label{eq4}
\end{eqnarray}
and applies the propagator $G(\omega)=\langle \!\langle A;A^\dagger
\rangle \!\rangle_\omega$. If we define the (orthogonal) operator $C$
as
\begin{equation}
[A,H]= \zeta A - i C, \qquad \langle \{C, A^\dagger\}_+\rangle =0, \label{eq5}
\end{equation}
we can express
\begin{eqnarray}
G(\omega) &=& G_0(\omega) + \frac{1}{\alpha^2} G_0(\omega)^2
\langle\!\langle C;C^\dagger \rangle\!\rangle_\omega, \nonumber \\ 
G_0(\omega)&=& \frac{\alpha}{\omega -\zeta}, \qquad \alpha=\langle
\{A,A^\dagger\}_+\rangle. \label{eq6}
\end{eqnarray}
Identifying the self energy $\Sigma(\omega)$ as the irreducible part
of $\langle\!\langle C;C^\dagger \rangle\!\rangle_\omega$ we can express
Eq.~(\ref{eq6}) as
\begin{equation}
G(\omega)={\alpha\over \omega - \zeta -\Sigma(\omega) },\qquad
\Sigma(\omega) \sim {1\over \alpha} \langle\!\langle C;C^\dagger
\rangle\!\rangle_{\omega}^{irr}. \label{eq7}
\end{equation}
Within the diagrammatic technique $\Sigma(\omega)$ corresponds to the
contribution of irreducible diagrams. Generally $\Sigma(\omega)$ can
be defined as a memory function within the Mori projection method
\cite{mori}.  Anyway, in most cases the successful application of the
method relies on the appropriate decoupling or other approximation of
the memory function $\Sigma(\omega)$ \cite{gotz}.

Applying the formalism to the propagator, Eq.~(\ref{eq3}), we have to deal
with the EQM for $\tilde c^{\dagger}_{{\bf k} s}$ with a nontrivial
normalization factor,
\begin{equation}
\alpha = \frac{1}{N}\sum_i \langle \{\tilde c_{i s},\tilde
 c^{\dagger}_{i s}\}_+ \rangle = 1 - \frac{c_e}{2} = \frac {1}{2}
 (1+c_h). \label{eq8}
\end{equation}
By taking explicitly into account the projection in Eq.~(\ref{eq2})
the EQM follow,
\begin{eqnarray}
[\tilde c_{is},H]&&= - \sum_j t_{ij}[(1-n_{i,-s})\tilde c_{js} +
S_i^{\mp} \tilde c_{j,-s}] \nonumber \\ +&&\frac{1}{4} J
\sum_{j~n.n. i} (2 s S_j^z \tilde c_{i s} + 2 S_j^{\mp} \tilde
c_{i,-s}- n_j \tilde c_{i s} )
\label{eq9}
\end{eqnarray}
with $s=\pm 1$. We express 'bosonic' variables in to terms of spin and
density operators, i.e., $n_{i,-s}=n_i/2 - sS_i^z$. Assuming as well
that we are dealing with a paramagnetic metal with $\langle {\bf S}_i
\rangle =0$ and a homogeneous electron density $\langle n_i \rangle =
c_e$, we get
\begin{equation}
-iC_{{\bf k} s}= [\tilde c_{{\bf k} s},H] - \zeta_{\bf k} \tilde
c_{{\bf k} s} , \label{eq10}
\end{equation}
and
\begin{eqnarray}
[\tilde c_{{\bf k} s},H]= &&[(1-\frac {c_e}{2}) \epsilon^0_{\bf k} -
J c_e]\tilde  c_{{\bf k} s} 
+ \frac{1}{\sqrt{N}}\sum_{\bf q} m_{\bf k q} \nonumber \\
&&\bigl[ s S^z_{\bf q} \tilde c_{{\bf k}-{\bf q},s}
 + S^{\mp}_{\bf q} \tilde c_{{\bf k}-{\bf q},-s} - 
\frac{1}{2} \tilde n_{\bf q} \tilde c_{{\bf k}-{\bf q}, s}\bigr],
\label{eq11}
\end{eqnarray}
where $\tilde n_i =n_i -c_e$, $m_{\bf k q}$ is the effective
spin-fermion coupling,
\begin{equation}
 m_{\bf k q}=2J \gamma_{\bf q} + \epsilon^0_{{\bf k}-{\bf q}}
\label{eq12}
\end{equation}
and $\epsilon^0_{\bf k}$ is the bare band dispersion, i.e., for the
model (\ref{eq1}) on a square lattice
\begin{eqnarray}
\epsilon^0_{\bf k}=-4t\gamma_{\bf k}-&& 4t'\gamma'_{\bf k},
\label{eq13} \\
\gamma_{\bf k}= \frac{1}{2}(\cos k_x +\cos k_y), \quad &&
\gamma_{\bf k}^\prime = \cos k_x \cos k_y.  \nonumber 
\end{eqnarray}
Equations (\ref{eq4}),(\ref{eq11}),(\ref{eq13}) define also the
'renormalized' band
\begin{eqnarray}
\zeta_{\bf k}&=& \frac{1}{\alpha} \langle \{[\tilde c_{{\bf k} s},H],
  \tilde c^{\dagger}_{{\bf k} s}\}_+\rangle \nonumber \\
&=& \bar \zeta - 4 \eta_1 t \gamma_{\bf k}  - 4 \eta_2 t' \gamma'_{\bf
k}, \label{eq14} \\
\eta_j &=& \alpha + \frac{1}{\alpha} \langle {\bf S}_0 \cdot {\bf S}_j \rangle,
\nonumber 
\end{eqnarray}
where $\eta_j$ are determined solely by short range spin correlations
and $\bar \zeta$ is a ${\bf k}$ independent term (still dependent on
various static correlations).

Above quantities determine the propagator
\begin{equation}
 G({\bf k},\omega)= \frac{\alpha}{\omega+\mu -\zeta_{\bf k} -
\Sigma({\bf k},\omega) }, \label{eq15}
\end{equation} 
and the corresponding spectral function $A({\bf k},\omega) =
-(1/\pi){\rm Im} G({\bf k},\omega)$, provided that we find a method to
evaluate $\Sigma({\bf k},\omega)$.

\section{Self energy}

\subsection{Undoped antiferromagnet}

It is desirable that in the case of an undoped AFM our treatment of
$\Sigma$ and the spectral function reproduces quite successful SCBA
equations \cite{kane} for the Green's function of a hole in an
AFM. Let us concentrate here on the relevant n.n. hopping, since the
$t'$ term represents a hopping on the same sublattice within an
ordered AFM and is therefore nearly free. For the SCBA the reference
state is the N\'eel state with $n_{is}=0,1$ for $i=A,B$ sublattices,
respectively, and the SCBA effective Hamiltonian can be written as
\begin{equation}
H_h= -t\sum_{\langle ij \rangle} (h_i h_j^\dagger a_j +
h_j h_i^\dagger a_j^\dagger) + H_J, \label{eq16}
\end{equation}
where $h_i$ represent holon operators and $a_i$ spin flip operators.
The corresponding holon EQM then follows from Eq.~(\ref{eq16})
\begin{equation}
-i \frac{d}{dt} h_i^\dagger = [h_i,H_h]= t \sum_{j~n.n.i}
h_j^\dagger (a_i^\dagger +a_j). \label{eq17}
\end{equation}
It is now straightforward to establish the relation of Eq.~(\ref{eq17})
with the EQM for $\tilde c_{is}$ by considering one
N\' eel sublattice $i=A$ with the reference state $n_{is}=1$. In this
case $1-n_{i,-s}=1$ and by formally replacing $\tilde c_{js}=\tilde
c_{j,-s} S^\mp_j $ we get by considering only the $t$ term in
Eq.~(\ref{eq9}),
\begin{equation}
i\frac{d}{dt} \tilde c_{is} \sim -t \sum_{j~ n.n.i} (S^\mp_i+ S^\mp_j)
\tilde c_{j,-s}. \label{eq18}
\end{equation}
To be consistent with the SCBA we neglect here the $J$ term in
Eq.~(\ref{eq9}) since anyhow $J\ll t$.  Within the linearized magnon
theory EQM (\ref{eq17}),(\ref{eq18}) are formally identical, so we can
furtheron follow the procedure of the evaluation of $\Sigma_{\rm
AFM}({\bf k}, \omega)$ within the SCBA to reproduce in the first place
spectral properties of an undoped AFM.  In this case we do not try to
improve the SCBA, since the latter approximation is simple and yields
both qualitatively and quantitatively good results consistent with
numerical studies and experiments. For an ordered 2D AFM where
relevant spin excitations are magnons with dispersion $\omega_{\bf q}$
we therefore get
\begin{eqnarray}
\Sigma_{\rm AFM}({\bf k},\omega)&=& \frac{1}{N} \sum_{\bf q}
M_{\bf kq}^2 G({\bf k}-{\bf q},\omega+\omega_{\bf q}), \label{eq19} \\
M_{\bf kq} &=& 4t (u_{\bf q} \gamma_{{\bf k}-{\bf q}}+v_{\bf q}
\gamma_{\bf k}),
\nonumber  
\end{eqnarray}
with
\begin{eqnarray}
u_{\bf q}=\sqrt{\frac{2J+\omega_{\bf q}}{2\omega_{\bf q}}},&& \qquad
v_{\bf q}=-\gamma_{\bf q} \sqrt{\frac{2J-\omega_{\bf q}}
{2\omega_{\bf q}}}, \nonumber \\
\omega_{\bf q}&=& 2J \sqrt{1-\gamma^2_{\bf q}}. \label{eq20} 
\end{eqnarray}
Since in a N\'eel state we have $\eta_1=0$ and hence the renormalized
band vanishes, i.e., $\zeta_{\bf k}=0$, we reproduce the usual SCBA
equations for the hole spectral function within the $t$-$J$ model.
The inclusion of the n.n.n. hopping $t'$ is also simple within the
SCBA since within the N\'eel state it does not induce a coupling to
spin flips in Eq.~(\ref{eq9}) and therefore enters $G({\bf
k},\omega)$, Eq.~(\ref{eq15}), only via the band term $\zeta_{\bf k}
\sim \bar \zeta -4t'\gamma'_{\bf k}$. It should also be noted that in
contrast to the usual SCBA our procedure deals directly with the
electron propagator and not with an unphysical holon one. Moreover it
allows a straightforward generalization to the case of finite doping.

\subsection{Short range transverse spin fluctuations}

For finite doping $c_h>0$ we assume that spin fluctuations remain
dominant at the AFM wavevector ${\bf Q}=(\pi,\pi)$ with the
characteristic inverse AFM correlation length $\kappa=1/\xi_{AFM}$.
The latter seems to be the case for BSCCO as well as
YB$_2$Cu$_3$O$_{6+x}$, but not for LSCO with pronounced stripe and
spin-density structures with ${\bf q}_{SDW} \ne{\bf Q}$. For the
former case one can divide the spin fluctuations into two regimes with
respect to $\tilde {\bf q}= {\bf q}-{\bf Q}$:

\noindent a) For $\tilde q>\kappa$ spin fluctuations are paramagnons,
i.e., they are propagating like magnons and are transverse to the
local AFM short-range spin ordering. Hence it make sense to use
Eqs.(\ref{eq19}),(\ref{eq20}) to represent paramagnon contribution to
the self energy restricting the sum to the regime $\tilde q>\kappa$.

\noindent b) For $\tilde q<\kappa$ spin fluctuations are essentially
not propagating modes but critically overdamped so deviations from the
long range order are essential.  Alternative approximations to
$\Sigma({\bf k},\omega)$ have to be used here, as discussed in
Sec.~III.C.

We should also take into account that the SCBA formalism has been
derived for an undoped AFM, i.e., for a hole spectral function at
$\omega<0$ where only (added) holes participate. Since we are
dealing with $c_h>0$ we take into account the scattering of hole-like
($\omega<0$) QP by replacing full propagator $G$ in Eq.~(\ref{eq19}) by
the hole part $G^-$,
\begin{equation}
G^\mp({\bf k},\omega)= \pm \int_{\mp \infty}^0
\frac{d\omega' A({\bf k},\omega')} {\omega-\omega'}, \label{eq21}
\end{equation}
However it is easy to see that an analogous contribution should arise
from the electron-like QP with $\omega>0$. At finite doping case we
therefore generalize (at $T=0$) Eq.~(\ref{eq19})
into the paramagnon contribution,
\begin{eqnarray}
\Sigma_{\rm pm}({\bf k},\omega)= &&\frac{1}{N} \sum_{q,\tilde q>
\kappa}  \bigl[ M_{\bf kq}^2  
G^-({\bf k}-{\bf q},\omega+\omega_{\bf q}) + \nonumber \\ && M_{{\bf
k}+{\bf q},{\bf q}}^2 G^+({\bf k}+{\bf q},\omega-\omega_{\bf q})],
\label{eq22}
\end{eqnarray}
which would emerge from Eq.~(\ref{eq18}). The consequence of
Eq.~(\ref{eq22}) is that in general Im$\Sigma_{\rm pm}({\bf
k},\omega>0) \neq 0$ so that also electron-like QP can be damped due
to magnon processes.  We do not consider here effects of $T>0$ which
could be easily incorporated through the magnon occupation, but in
most cases do not have a strong influence at low $T<J$.

Here we stress two features of our approximation for paramagnon
contribution $\Sigma_{\rm pm}$:

\noindent a) we are dealing with a strong coupling theory due to 
$t > \omega_{\bf q}$ and a selfconsistent
calculation of $\Sigma_{\rm pm}$ is required,

\noindent b) resulting $\Sigma_{\rm pm}({\bf k},\omega)$ as well
$A({\bf k},\omega)$ as are at low doping quite asymmetric with respect
to $\omega = 0$.  As in an undoped AFM the hole part $G^-$ with the
weight $\propto (1-c_h)/2 \sim 1/2$ generates a large incoherent part
in $A({\bf k},\omega \ll 0)$. On the other hand, $G^+$ has less weight
$\propto c_h$ and consequently the scattering of electron QP is in
general much less effective.

\subsection{Coupling to longitudinal spin fluctuations}

Discussing the self energy at finite doping, Eq.~(\ref{eq22})
represents only one contribution and we have to reconsider EQM
(\ref{eq10}),(\ref{eq11}). We note that at $c_h>0$ $C_{{\bf k}s}$
contains a remainder of a 'free' term $\propto \tilde c_{{\bf k}s}$,
which should be however neglected when evaluating the 'irreducible'
part entering $\Sigma$, Eq.~(\ref{eq7}). Considering within the
simplest approximation only the mode-coupling terms in
Eq.~(\ref{eq11}) we also neglect the coupling to density fluctuation
$\tilde n_{\bf q}$ which should contribute much less to $\Sigma$ in
the absence of charge ordering or charge instabilities at low doping.

Taking into account only spin fluctuations, at $c_h>0$ we are dealing
with a paramagnet without an AFM long-range order and besides the
paramagnon excitations also the coupling to longitudinal spin
fluctuations become crucial. The latter restore the spin rotation
symmetry in a paramagnet and EQM (\ref{eq11}) naturally introduce
such a spin-symmetric coupling.  Assuming within a simplest
approximation that the dynamics of fermions and spins is independent,
\begin{eqnarray}
\langle S^z_{\bf q}(t)\tilde c_{{\bf k}-{\bf q},s}(t)
S^z_{-{\bf q}'} \tilde c^{\dagger}_{{\bf k}-{\bf q}',s} \rangle
\sim \nonumber \\
\delta_{{\bf q} {\bf q}'} \langle S^z_{\bf q}(t) 
S^z_{-{\bf q}} \rangle \langle \tilde c_{{\bf k}-{\bf q}, s}(t)
\tilde c^{\dagger}_{{\bf k}-{\bf q}, s} \rangle, \label{eq23}
\end{eqnarray}
we  get for the contribution of longitudinal fluctuations
\begin{eqnarray}
\Sigma_{\rm lf}({\bf k},\omega) &=&\frac{r_s}{\alpha} \sum_{\bf q} 
\tilde m^2_{\bf k q}
\int \int \frac{d\omega_1 d\omega_2}{\pi} g(\omega_1,\omega_2)
\nonumber \\
&&\frac{\tilde A({{\bf k}-{\bf q}},\omega_1) \chi''({\bf q},\omega_2)}
{\omega-\omega_1-\omega_2 }, \\ g(\omega_1,\omega_2)&=& f(-\omega_1)
+ \bar n(\omega_2) = \frac{1}{2}\bigl [{\rm th}\frac{\beta\omega_1}{2}+{\rm
cth}\frac{\beta\omega_2}{2} \bigr], \label{eq24} \nonumber  
\end{eqnarray}
where $\chi$ is the dynamical spin susceptibility
\begin{equation}
\chi({\bf q},\omega)=-i\int_0^{\infty} {\rm e}^{i \omega t} \langle
[S^z_{\bf q}(t) , S^z_{-\bf q}] \rangle d t. \label{eq25}
\end{equation}
Such an approximation for $\Sigma$ has been introduced within the
$t$-$J$ model in Ref.~\cite{prel}. However, quite analogous treatment
has been employed previously in the Hubbard model \cite{kamp} and more
recently within the spin-fermion model \cite{chub,schm}.

Several comments are in order to define quantities entering Eq.~(\ref{eq24}): 

\noindent a) EQM (\ref{eq11}) induce an effective spin-fermion
coupling, which would emerge also from a phenomenological spin-fermion
Hamiltonian with the coupling parameter $m_{\bf kq}$,
Eq.~(\ref{eq12}).  In order that such a Hamiltonian is hermitian, the
coupling should satisfy
\begin{equation}
\tilde m_{{\bf k},{\bf q}} = \tilde m_{{\bf k}-{\bf q},-{\bf q}},
\label{eq26}
\end{equation}
which is in general not the case with the form Eq.~(\ref{eq12}), therefore
we use furtheron instead the symmetrized coupling
\begin{equation}
\tilde m_{\bf kq}= 2J \gamma_{\bf q}+ \frac{1}{2} 
(\epsilon^0_{{\bf k}-{\bf q}}+\epsilon^0_{\bf k}). \label{eq27}
\end{equation}
Here we should point out that in contrast to previous related studies
of phenomenological spin-fermion coupling \cite{kamp,chub,schm}, our
$\tilde m_{\bf kq}$ (as well as $m_{\bf kq}$) is strongly dependent on
both ${\bf q}$ and ${\bf k}$. It is essential that in the most
sensitive parts of the FS, i.e., along the AFM zone boundary ('hot'
spots) where $k=|{\bf Q}-{\bf k}|$, the coupling is in fact quite
modest determined solely by $J$ and $t'$.

\noindent b) Since we are dealing with the paramagnetic state, all
quantities should be spin invariant, i.e., the $\chi^{\alpha\beta}({\bf
q},\omega) =\delta_{\alpha\beta} \chi({\bf q},\omega)$. Since EQM
(\ref{eq11}) are invariant under spin rotations we have besides the
$S^z$ term analogous terms with $S^-, S^+$. Still we expect $r_s=1$
instead of $r_s=3$ since only the coupling to longitudinal (to local
N\' eel spin order) spin fluctuations is considered here, while the
coupling to transverse fluctuations has been already taken into
account by $\Sigma_{\rm pm}$.

\noindent c) In $\Sigma_{\rm lf}$ only the part corresponding to
irreducible diagrams should enter, so there are restrictions on the
proper decoupling. We will be interested mostly in the situation with
a pronounced AFM short-range order where longitudinal fluctuations are
slow, i.e., with the characteristic frequencies $\omega_\kappa \alt 2J
\kappa \ll J$. The regime is close to that of quasistatic $\chi({\bf
q},\omega)$ where the simplest and also quite satisfactory
approximation is to insert for $\tilde A$ the unrenormalized $A^0$,
the latter corresponding in our case to the spectral function without
$\Sigma_{\rm lf}$ but with $\Sigma=\Sigma_{\rm pm}$.  Such an
approximation has been introduced in the theory of a pseudogap in CDW
systems \cite{lee}, used also in related works analyzing the role spin
fluctuations \cite{kamp},\cite{chub}, and recently more extensively
examined in Ref.\cite{mill}.  In the opposite case of full
self-consistent treatment with $\tilde A=A$ we would overcount the
influence of fluctuations, although the results would probably appear
not so much different as shown on simpler systems \cite{mill}.

For $\chi({\bf q},\omega)$, Eq.~(\ref{eq25}), at $c_h>0$ and possibly
$T>0$ we do not have a corresponding theory, so we treat it as an
input.  On one hand, $\chi({\bf q},\omega)$ is restricted by the sum
rule,
\begin{equation}
{1\over N} \sum_{\bf q} \int_0^{\infty} {\rm
cth}(\frac{\beta\omega}{2}) \chi''({\bf q},\omega) d \omega =
\frac{\pi} {4} (1-c_h). \label{eq28}
\end{equation}
At the same time, the system is close to the AFM instability, so we
assume spin fluctuations of the overdamped form \cite{mont}
\begin{equation}
\chi''({\bf q},\omega) \propto\frac{ \omega}
{(\tilde q^2 +\kappa^2) (\omega^2+\omega_\kappa^2)}. \label{eq29}
\end{equation}
The appearance of the pseudogap and the form of the FS are not
strongly sensitive to the particular form of $\chi''({\bf q},\omega)$
(at given characteristic $\kappa$ and $\omega_\kappa$) provided that
$\chi({\bf q},\omega)$ is not singular as, e.g., is the case of
marginal Fermi liquid scenario \cite{varm}. It has been shown
\cite{prel} that the latter form is needed to get generally overdamped
QP with vanishing QP weight (at $T=0$) in spectral functions at
intermediate doping.

\section{Pseudogap analysis}

Full calculation of the spectral functions $A({\bf k},\omega)$ within
the presented theory requires a selfconsistent solution for
$\Sigma=\Sigma_{\rm pm}+\Sigma_{\rm lf}$, where besides the model
parameters $t,t',J$ and the doping $c_h$, an input are also $\mu,
\kappa, \eta_1, \eta_2$. $\kappa, \eta_1, \eta_2$ are given by
short-range spin correlations dependent mostly on $c_h$ and can be
taken from various analytical \cite{sing} and numerical
\cite{bonc,dago} calculations within the $t$-$J$ model. At the same
time in a selfconsistent theory $\mu$ should be fixed via the DOS
\begin{equation}
{\cal N}(\omega)={2\over N} \sum_{\bf k}  A({\bf k}, \omega). \label{eq30}
\end{equation}
as
\begin{equation}
c_h= 1 - \int_{-\infty}^\infty f(\omega) {\cal N}(\omega)d\omega. \label{eq31}
\end{equation}
Results of such selfconsistent calculation are presented in Sec.~V.

On the other hand, to establish characteristic features of the
pseudogap and the development of the FS we follow first a simplified
analysis. We namely notice that the effect of $\Sigma_{\rm pm}$ are
threefold: 

\noindent a) to induce a large incoherent component in the spectral
functions at $\omega \ll 0$ in particular at low and intermediate
doping,

\noindent b) to renormalize the effective QP band relevant to the
behavior at $\omega \sim 0$ and at the FS, and

\noindent c) to cause a transition of a large FS into a small
hole-pocket-like  FS at $c_h<c_h^* \ll 1$,

Result b) can serve as a starting point for the discussion of the
pseudogap and FS features at finite doping.  If we define the
effective band as
\begin{eqnarray}
\epsilon_{\bf k}^{\rm ef}&=& Z^{\rm ef}_{\bf k} [\zeta_{\bf k} + 
\Sigma_{\rm pm}({\bf k},0) -\mu], \nonumber \\ 
Z^{\rm ef}_{\bf k}&=&\Bigl[ 1- \frac{\partial \Sigma_{\rm pm}
({\bf k},\omega)}{\partial \omega}\Bigr|_{\omega=0}\Bigr]^{-1},
\label{eq32}
\end{eqnarray}
we get for the effective spectral function 
\begin{equation}
A^0_{\rm ef}({\bf k},\omega)=\alpha Z^{\rm ef}_{\bf k} \delta(\omega +\mu
-\epsilon_{\bf k}^{\rm ef}), \label{eq33}
\end{equation}
which can be used to evaluate $\Sigma_{\rm lf}$. We restrict
ourselves here to the regime of intermediate (not too small) doping,
where $\epsilon_{\bf k}^{\rm ef}$ defines the large FS. 

Let us concentrate on results for $T=0$. The simplest situation where
$\Sigma_{\rm lf}$ can be evaluated analytically is the quasi-static and
single-mode approximation (QSA) which is meaningful if $\omega_\kappa
\ll t, \kappa \ll 1$. In this case we insert into Eq.~(\ref{eq24})
\begin{equation}
\frac{1}{\pi} \chi''({\bf q},\omega) \sim \frac{1}{4} 
\delta({\bf q}-{\bf Q}) [\delta(\omega-\nu) -
\delta(\omega+\nu)], \label{eq34}
\end{equation}
with $\nu \to 0$. We get
\begin{equation}
\Sigma^{QSA}_{\rm lf}({\bf k},\omega)= \frac{r_s m^2_{\bf kQ}}{4} 
\frac{Z^{\rm ef}_{{\bf k}-{\bf Q}}} 
{\omega - \epsilon^{\rm ef}_{{\bf k}-{\bf Q}}} \label{eq35}
\end{equation}
and
\begin{eqnarray} 
G^{QSA}({\bf k},\omega)&=& \frac{\alpha Z^{\rm ef}_{\bf k} (\omega -
\epsilon^{\em ef}_{{\bf k}-{\bf Q}})} {(\omega - \epsilon^{\rm ef}_{{\bf
k}-{\bf Q}})(\omega - \epsilon^{\rm ef}_{\bf k}) - \Delta^2_{{\bf k}}
},\label{eq36} \\ 
\Delta^2_{{\bf k}} &=&\frac{r_s}{4} Z^{\rm ef}_{\bf k} 
Z^{\rm ef}_{{\bf k}-{\bf Q}}
m^2_{\bf kQ}, \nonumber 
\end{eqnarray} 
The spectral functions show in this approximation two branches of
$E^\pm$, separated by the gap which opens along the AFM zone boundary
${\bf k}={\bf k}_{AFM}$ where $\epsilon^{\rm ef}_{{\bf k}-{\bf
Q}}=\epsilon^{\rm ef}_{{\bf k}}$.  Since $\gamma_{{\bf k}_{AFM}}=0$ the
relevant (pseudo)gap scale is
\begin{equation}
\Delta^{PG}_{\bf k} = |\Delta_{{\bf k}_{AFM}}| = 
\frac{Z^{\rm ef}_{\bf k}}{2} 
\sqrt{ r_s} |2J - 4 t' {\rm cos}^2 k_x|. \label{eq37}
\end{equation}
It is instructive to realize that $\Delta^{PG}_{\bf k}$ does not
depend on $t$, but rather on smaller $J$ and in particular $t'$. For
$t'<0$ the gap is largest at $(\pi,0)$, consistent with
experiments. Whether the (pseudo)gap appears at the Fermi energy
$\omega=0$ depends, however, on properties of $\epsilon^{\rm ef}_{{\bf
k}_{AFM}}$. We do not expect that the gap opens along the whole AFM zone
boundary, since in most cases $\epsilon^{\rm ef}_{{\bf k}_{AFM}}$ crosses
zero along $(\pi/2,\pi/2)$-$(\pi,0)$ so that within the QSA $E^-_{\bf
k}$ forms a hole-pocket-like FS. In fact, the results of the QSA are
equivalent to the system with long range spin-density-wave order
(AFM), where the doubling of the unit cell appears. 

Within the simplified effective band approach, Eq.~(\ref{eq31}), it is
not difficult to evaluate numerically $\Sigma_{\rm lf}$ beyond the
QSA, by taking explicitly $\chi({\bf q},\omega)$, Eq.~(\ref{eq29}), for
$\kappa>0$ and $\omega_\kappa=2J\kappa$. Integrals in
Eq.~(\ref{eq24}) can be performed mostly analytically if we linearize the
dependence of $\epsilon^{\rm ef}_{\bf k}$ within the relevant interval
$\delta k \alt \kappa$.

Let us for illustration present in this Section results characteristic
for the development of spectral functions with most sensitive
parameters $\kappa$ and $\mu$, which both simulate the variation with
doping.  We fix furtheron the model parameter $J/t=0.3$ as relevant
for cuprates. We take here $t'/t=-0.3$ close to values quoted for
BSCCO. For simplicity we assume first that the effective band
$\epsilon^{\rm ef}_{\bf k}$ is just renormalized $\epsilon_{\bf k}$
(justified for an intermediate doping, see Sec.~V) with fixed values
$t_{\rm ef}/t=0.3$, $t'_{\rm ef}/t=-0.1$ and $Z^{\rm ef}=0.4$. More
realistic treatment would require the variation of latter parameters
with $c_h$ but results remain qualitatively similar. For $\kappa$ we
take in accord with experiments \cite{imad} and numerical results on
the $t$-$J$ model \cite{bonc,dago,sing} $\kappa \sim \sqrt{c_h}$.

The choice of $\mu$ is somewhat more arbitrary since within an
effective band approach the sum rule, Eq.~(\ref{eq30}), cannot be used
as a criterion. Nevertheless it is evident that $\mu$ determines the
shape and the volume of the FS. In the following examples we choose
$\mu$ such that at given $\kappa$ the DOS at the Fermi energy, ${\cal
N}(0)$, reaches a local minimum. This means that effectively the
states near $(\pi,0)$ are in the pseudogap and that the truncation of
the FS is most pronounced (at given $\kappa$). Such a choice of $\mu$
in fact also yields the volume of the FS (except at extreme $\kappa
\ll 1$) quite close to the one consistent with the Luttinger theorem
\cite{lutt}.

In Fig.~1 we first present results for $A({\bf k},\omega=0)$ at $T=0$
for a broad range of $\kappa=0.01 - 0.6$. Curves (evaluated at small
additional smearing $\epsilon = 0.02 t$) in fact display the effective
FS determined by the condition $G^{-1}({\bf k}_F,0)=0$. At the same
time, intensities $A({\bf k},\omega=0)$ correspond to the
renormalization factor $Z_F$. We can comment the development as
follows. At extremely small $\kappa=0.01$ we see the hole-pocket FS
which follows from the QSA in Eq.~(\ref{eq36}). In spite of small
$\kappa$ the 'shadow' side of the hole pocket has smaller
$Z_F$. Already small $\kappa \sim 0.05$ destroys the 'shadow' side of
the pocket, i.e., the solution $G^{-1}=0$ on the latter side
disappears since the singularity in $\Sigma_{\rm lf}$,
Eq.~(\ref{eq35}), is smeared out by finite $\kappa$. On the other
hand, in the gap emerge now QP solutions with very weak $Z_F \ll 1$
which reconnect the FS into a large one. We are dealing nevertheless
with effectively truncated FS with well developed arcs. The effect of
larger $\kappa$ is essentially to increase $Z_F$ in the gaped region,
in particular near $(\pi,0)$. Finally, for large $\kappa = 0.6$ which
corresponds to the regime consistent with optimal doping or overdoping
in cuprates, $Z_F$ is essentially only weakly decreasing towards
$(\pi,0)$ and the FS is well pronounced and concave as naturally
expected for $t'<0$.

\noindent
\begin{figure}[htb]
\center{\epsfig{file=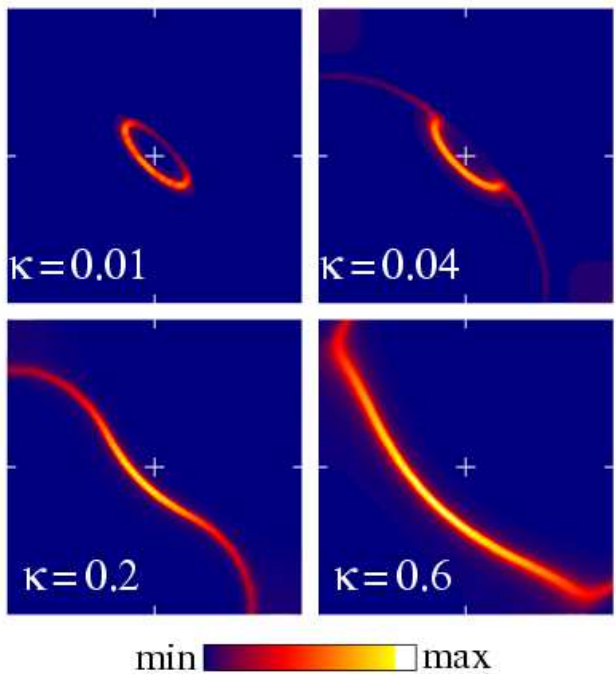,width=85mm,angle=-0,clip=}}\\
\caption{(color) Contour plot of spectral functions $A({\bf k},\omega=0)$ at
$T=0$ for various $\kappa$ in one quarter of the Brillouin zone.}
\end{figure}  

In order to understand the pseudogap features at low but finite
$\kappa$ we present in Figs.~2,3 $A({\bf k},\omega)$ for $\kappa=0.1$.
Spectra in Fig.~3 are presented along the lines a - d in the Brillouin
zone as shown in Fig.~2. As expected from Eq.~(\ref{eq36}) the
pseudogap is smallest along the zone diagonal (line a) where,
moreover, the pseudogap appears at $\omega>0$, so that it would not be
seen in ARPES.  Lines a and b are thus examples of the region where
arcs of the FS are well pronounced, i.e., their QP weight is not
strongly renormalized, $Z_F \alt Z^{\rm ef}$. On the other hand,
following lines c and d the chemical potential $\omega=0$ falls into
the pseudogap. We see in Fig.~3(a),(b) that QP in fact crosses
coherently the FS ($\omega=0$) although with very small $Z_F \ll
1$. It is evident from Fig.~1 that $Z_F$ within the pseudogap remains
small only for $\kappa \ll 1$ while it increases and finally smears
out the concept of the pseudogap for $\kappa \agt \kappa^* \sim 0.5$.

\begin{figure}[htb]
\center{\epsfig{file=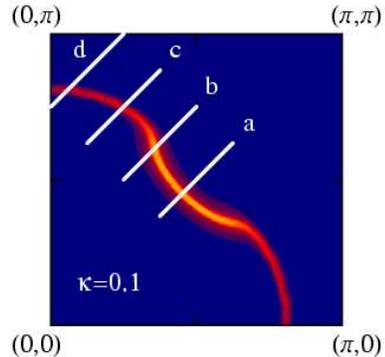,width=50mm,angle=-0,clip=}}\\
\caption{(color) $A({\bf k},\omega=0)$ for $\kappa=0.1$
and lines a) - d) used in Fig.~3.}
\end{figure}     

\vskip -.5cm
\noindent 
\begin{figure}
\center{\epsfig{file=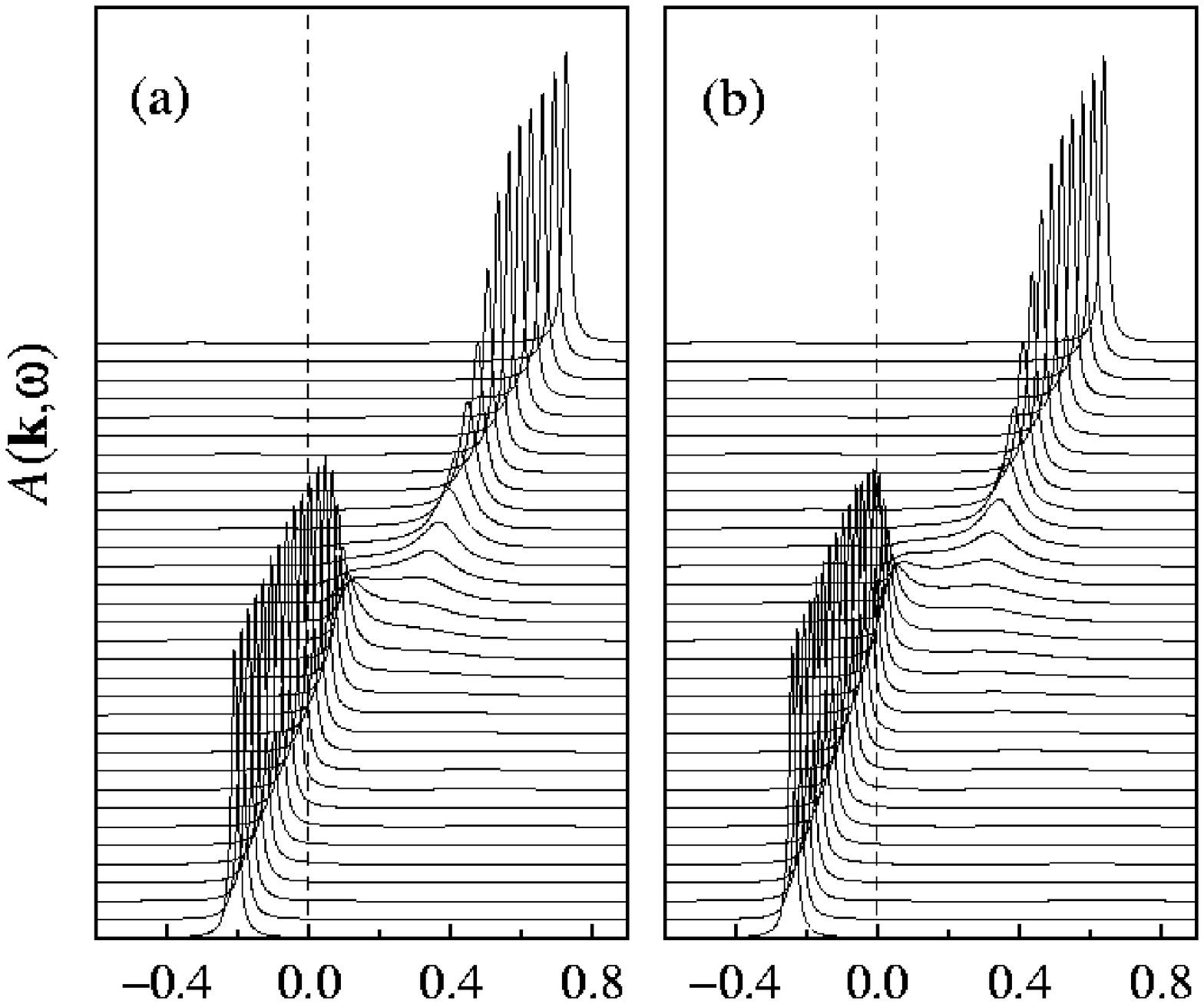,width=57mm,angle=-90,clip=}}\\[-5mm]
\vskip -5 mm
\center{\epsfig{file=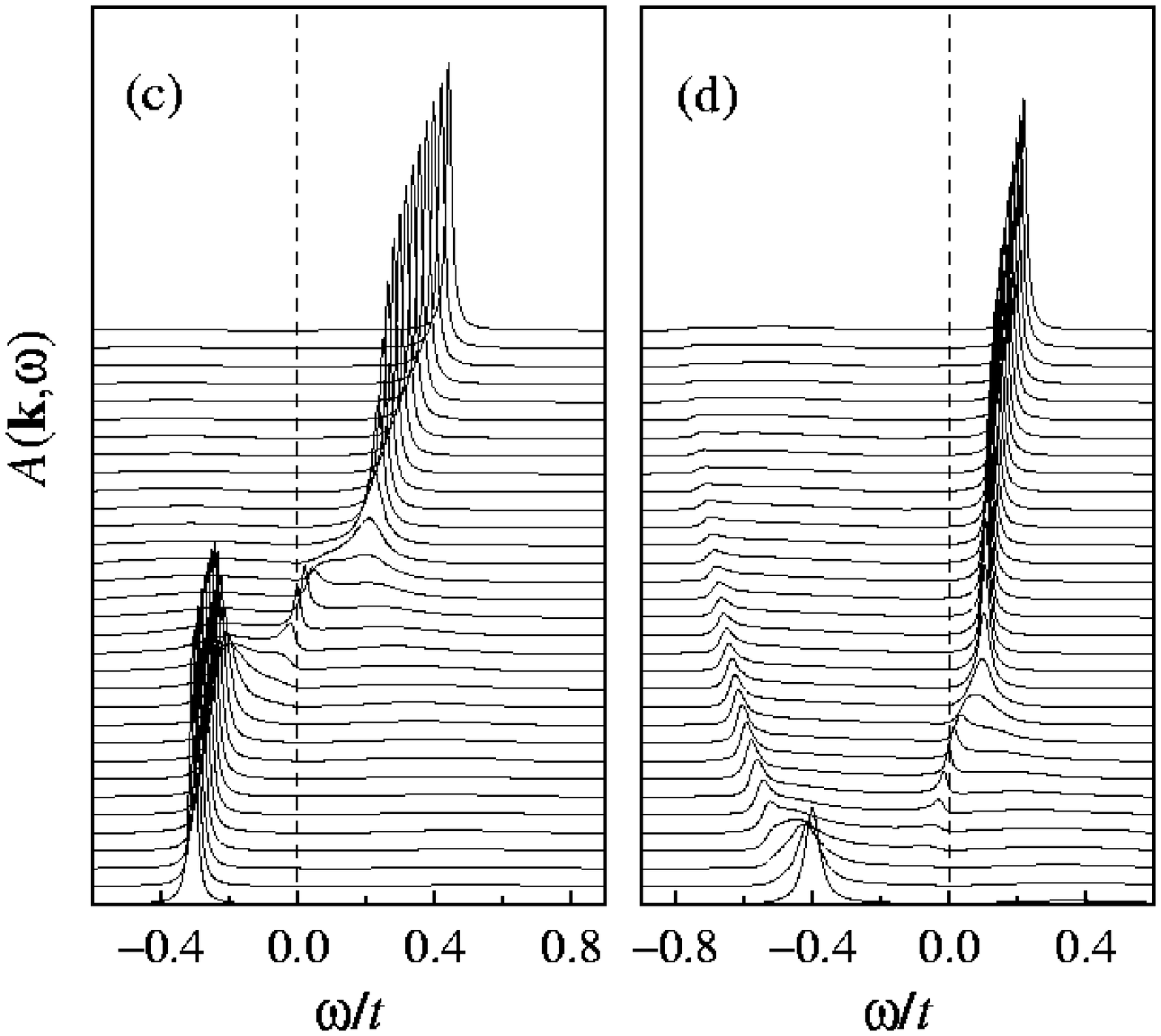,width=58mm,angle=-90,clip=}}\\
\caption{$A({\bf k},\omega)$ for $\kappa=0.1$ along different
directions in the Brillouin zone, corresponding to Fig.~2.}
\end{figure}

It is quite remarkable to notice that in spite of $Z_F \ll 1$ the QP
velocity $v_F$ is not diminished within the pseudogap. In fact it is
even enhanced, as seen in Fig.~3(c) and even more clearly in Fig.~4
where the contour plot of $A({\bf k},\omega)$ is shown corresponding
(with restricted $k$ span) to Fig.~3(c). Again, it is well evident in
Fig.~4 that QP is well defined at the FS, while it becomes fuzzy at
$\omega \neq 0$ merging with the solutions $E_{\bf k}^\pm$,
respectively, away from the FS.

\noindent
\begin{figure}[htb]
\center{\epsfig{file=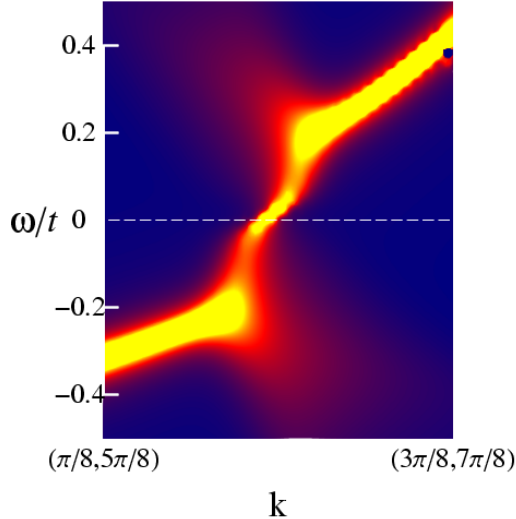,width=70mm,angle=-0,clip=}}\\
\caption{(color) Contour plot of $A({\bf k},\omega)$ for $\kappa=0.1$,
corresponding to Fig.~3(c). Note large quasiparticle velocity around
$\omega\sim0$.}
\end{figure}

Presented formalism offers a possible scenario for the evolution of
the FS with doping from a pocket-like into a large one. In order to
explain results in Figs.~(3),(4) on the effective truncation of the FS and
the character of QP within the pseudogap we note that it is
essentially enough that both $\kappa$ and $\omega_\kappa$ are finite
to yield well defined FS. Since gross features do not depend on the
particular form Eq.~(\ref{eq29}) we present here simplified analysis using
instead
\begin{equation}
\chi^{\prime\prime}({\bf Q}+\tilde {\bf q},\omega)=\left\{
\begin{array}{ll} C
[\delta(\omega-\omega_\kappa)-\delta(\omega+\omega_\kappa)], \tilde
q_\perp < \kappa, \\ 0,\qquad\tilde q_\perp > \kappa,
\end{array} \right.\label{eq38}
\end{equation}
where $\tilde q_\perp$ denotes the component perpendicular to the AFM
zone boundary.  Let us assume that $\epsilon_{\bf k}^{\rm ef}-\mu=\epsilon
\sim 0$ and $\epsilon_{{\bf k}-{\bf Q}}^{\rm ef}-\mu =\bar \epsilon \sim
0$. We also linearize dispersion $\epsilon_{\bf k}^{\rm ef}$ at the FS and
take that ${\bf v}^{\rm ef}_{\bf k} \parallel (1,1)$, so that we get from
Eq.~(\ref{eq24}),
\begin{equation}
\Sigma(\epsilon,\omega)=-\frac{\Delta^2}{2 w}\log
\frac{(w+\omega_\kappa +\bar \epsilon-\omega)(\omega_\kappa +\omega)}
{(w+\omega_\kappa -\bar \epsilon+\omega)(\omega_\kappa -\omega)}
\label{eq39}
\end{equation}
where $w=v_{\bf k}^{\rm ef} \kappa$ and $\Delta=\Delta_{\bf k}$. Let us
evaluate QP properties on the FS assuming that it is located at
$\bar \epsilon=0$, i.e., on the AFM zone boundary. We obtain
from Eq.~(\ref{eq39}) the QP weight  $Z_F$
\begin{equation}
\frac{Z_F}{Z^{\rm ef}} = \Bigl [ 1-\frac{\partial\Sigma'}{\partial 
\omega}\Bigr|_{\omega=0,\bar \epsilon=0} \Bigr ]^{-1}=
\Bigl [1 + \frac{\Delta^2}{\omega_\kappa(\omega_\kappa+w)}
\Bigr ]^{-1}. \label{eq40}
\end{equation}
This clearly leads to and explains $Z_F \ll 1$ for $\omega_\kappa
w \sim 2 v_{\bf k}^{\rm ef} J \kappa^2 \ll \Delta^2$. This is generally
the case within the gaped part of the FS for small $\kappa<\kappa^*$, as
shown in Fig.~1. It should be also noted that the latter condition is
essentially always satisfied near $(\pi,0)$ where $v_{\bf k}^{\rm ef}
\sim 0$ and consequently also $w \sim 0$.

Let us evaluate in the same way the QP renormalized velocity $v_F$ 
at the FS. Here we realize that the ${\bf k}$ dependence of
$\Sigma'$ is essential. The latter is in Eq.~(\ref{eq40}) given by the
$\epsilon$ dependence,
\begin{eqnarray}
&&\frac{v_F}{v^{\rm ef}_{\bf k}}= (1+ \frac{\partial\Sigma'}
{\partial \epsilon} )\frac{Z_F}{Z^{\rm ef}}, \nonumber \\
&&\frac{\partial\Sigma'}{\partial \epsilon}\Bigr|_{\omega=0,\bar \epsilon=0}=
\frac{\Delta^2}{\omega_\kappa(\omega_\kappa+w)}, \label{eq41}
\end{eqnarray}
which in contrast to $Z_F$ leads to an enhancement of $v_F$. 
In the case $\omega_\kappa w \ll \Delta^2$ we thus get
\begin{equation}
\frac{v_F}{v^{\rm ef}_{\bf k}} \sim \frac{\omega_\kappa}{w} \sim
\frac{2J}{v_{\bf k}}. \label{eq42}
\end{equation}
Final $v_F$ is therefore not strongly renormalized, since $2J$ and
$v^{\rm ef}_{\bf k}$ are of similar order. Furthermore, $\tilde v_F$
is enhanced in the parts of FS where $v^{\rm ef}_{\bf k}$ is small, in
particular near $(\pi,0)$ point. The situation is thus very different
from 'local' theories where $\Sigma({\bf k},\omega) \sim
\Sigma(\omega)$ and the QP renormalization is governed only by
$Z_F$. In our case the 'nonlocal' character of $\Sigma({\bf
k},\omega)$ is essential in order to properly describe QP within the
pseudogap region.

Let us discuss further the behavior of the DOS $\cal N(\omega)$,
Eq.~(\ref{eq30}). It is evident from Fig.~1 that the contribution to
${\cal N}(\omega \sim 0)$ will come mostly from FS arcs near the zone
diagonal while the gaped regions near $(\pi,0)$ will contribute less.
Results in Fig.~5 (full lines) the development of ${\cal N}(\omega)$
with $\kappa$, as corresponding to FS in Fig.~1. We see that the DOS
indeed reveal a pseudogap at $\omega < \Delta$ however the pseudogap
is visible only for $\kappa < 0.5$ and deepens for $\kappa \to 0$.

\vskip -2 cm
\noindent
\begin{figure}[htb]
\center{\epsfig{file=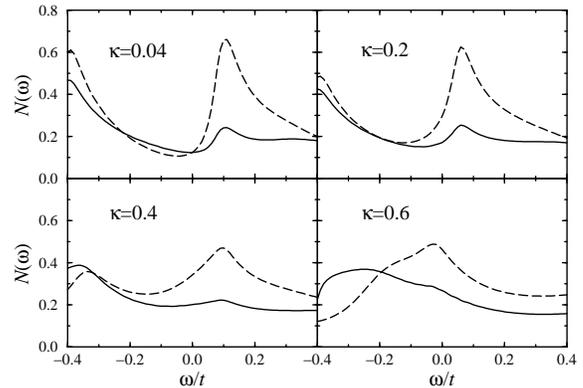,width=60mm,angle=-90,clip=}}\\
\caption{Density of states ${\cal N}(\omega)$ (full lines) and
weighted DOS ${\cal N}_w(\omega)$ (dashed lines) for different
$\kappa$.}
\end{figure}

The DOS is measured in cuprates via angle integrated PES, e.g. for LSCO in 
\cite{ino}, as well as via the scanning tunneling microscopy (STM)
\cite{renn}. It is well possible that within both experiments the matrix
elements are essential leading to enhanced contribution near the x
points in the Brillouin zone. It has been proposed that for the
$c$-axis conductivity \cite{ioff} the interplanar hopping should be
weighted by the matrix element
\begin{equation}
w({\bf k})= (\cos k_x - \cos k_y)^2. \label{eq43}
\end{equation}
The same arguments as for the $c$-axis conductivity might apply also
for the STM effective DOS as well as for integrated PES, therefore we
present also the weighted DOS ${\cal N}_w$ where $w({\bf k})$ is
introduced additionally into Eq.~(\ref{eq30}). Results also presented
in Fig.~5 (dashed line) show much stronger pronounced pseudogap, in
particular at low $\kappa$. This is quite evident since $w({\bf k})$
essentially destroys the effect of FS arcs near $(\pi/2,\pi/2)$
which present the main contribution (due to small velocity in
hole-pocket FS) to the usual ${\cal N}(\omega)$.

In Fig.~6(a) we show the average $Z_{\rm av}$ along the FS, as well as the
QP DOS, defined as
\begin{equation}   
{\cal N}_{QP}= \frac{\alpha Z^{\rm ef}}{2\pi^2} \oint 
\frac{dS_F}{v({\bf k})}. \label{eq44}
\end{equation}   
in Fig.~6(b) we present as well as the dependence of DOS at the FS,
both ${\cal N}(0)$ and ${\cal N}_w(0)$, as a function of $\kappa$.
Note that ${\cal N}_{QP}$ should be relevant for the specific heat,
i.e., ${\cal N}_{QP} \propto \gamma = C_V/T$ at low $T$ (provided that
we are dealing with a normal Fermi liquid). It is quite important to
understand that decreasing $\kappa$ (smaller doping) also means the
decrease of ${\cal N}_{QP}$, which is consistent with the observation
of the pseudogap also in the specific heat in cuprates \cite{lora}. We
note here that such a behavior is not evident when one discusses the
metal-insulator transition. Namely, in a Fermi liquid with (nearly)
constant Fermi surface one can drive the metal-insulator transition by
$Z_{\rm av} \to 0$ and within an assumption of a local character
$\Sigma(\omega)$ this would lead to $v_F \to 0$ and consequently to
${\cal N}_{QP} \to \infty$. Clearly, the essential difference in our
case is that within the pseudogap regime $\Sigma({\bf k},\omega)$ is
nonlocal, allowing for the simultaneous decrease of ${\cal N}(0)$ and
${\cal N}_{QP}$.

Finally, let us comment on the influence of finite $T$. While $T$
enters within this approach also via effective parameters as
$v^{\rm ef}_{\bf k}$ and predominantly $\kappa(T)$, here we consider only
the direct effect via the thermodynamic factor in Eq.~(\ref{eq24}). It is
evident that $T>0$ smears out $\Sigma_{\rm lf}$. 
This becomes important at
small $\kappa$ in particular for QP in the pseudogap regime. In Fig.~7
we present $A({\bf k},\omega)$, corresponding to Fig.~3(c), for several
values of low $T$. The main conclusion is, that weak (but sharp) QP peak with
$Z_F \ll 1$ at $T=0$ is smeared out already by very small $T>T^s \sim
0.02~t$ and is not at all visible (there is no overdamped peak) at
higher $T$, the remainder being an incoherent background at $\omega
\sim 0$ for $T>T^s$. This is important to realize that ARPES
experiments in fact do not observe no well defined QP peak near $(\pi,0)$
in the underdoped regime at $T>T_c$. 
\vskip -1cm
\noindent
\begin{figure}[t]
\center{\epsfig{file=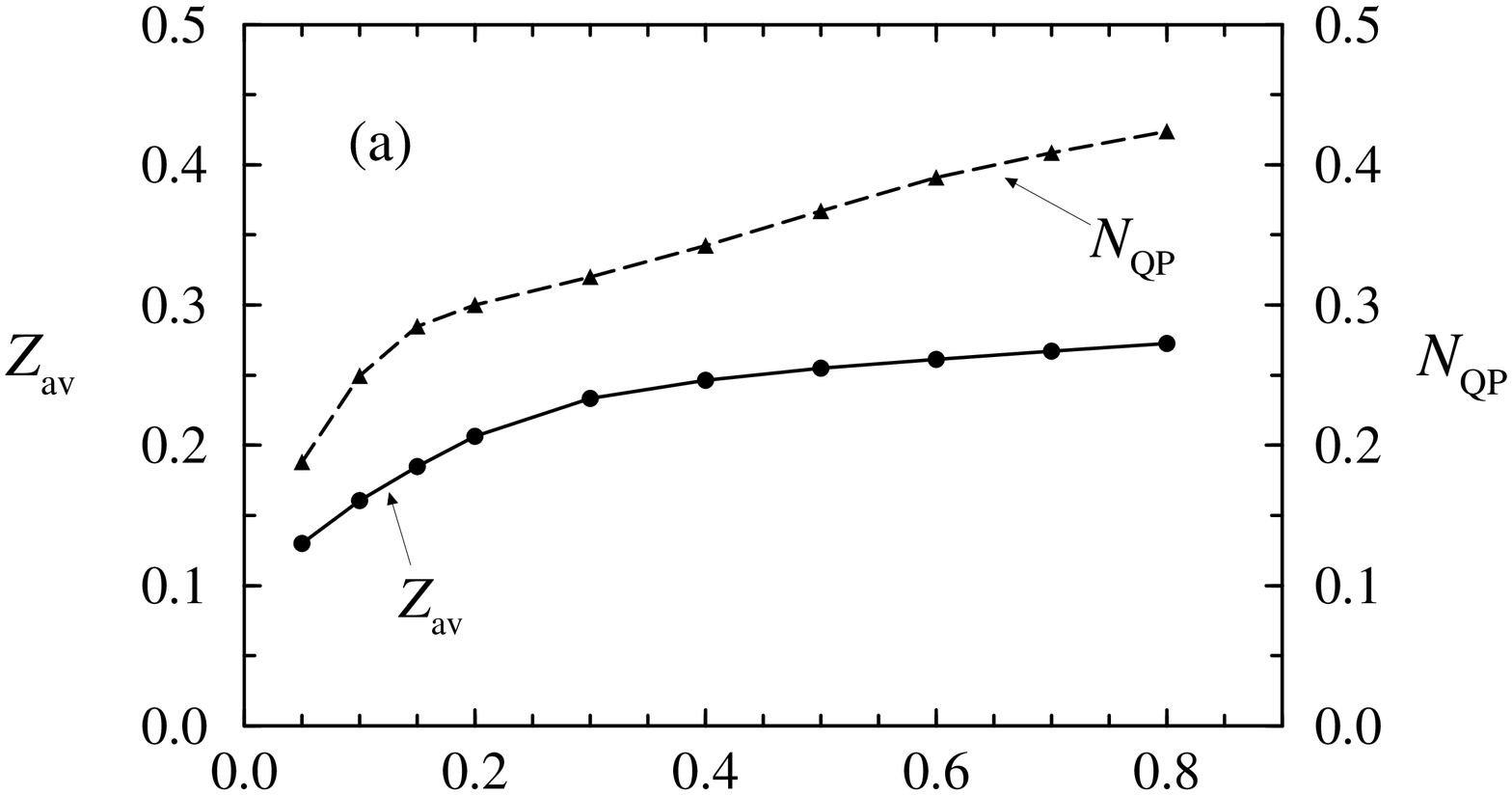,width=65mm,angle=-90,clip=}}\\[-25 mm]
\vskip -25 mm 
\center{\epsfig{file=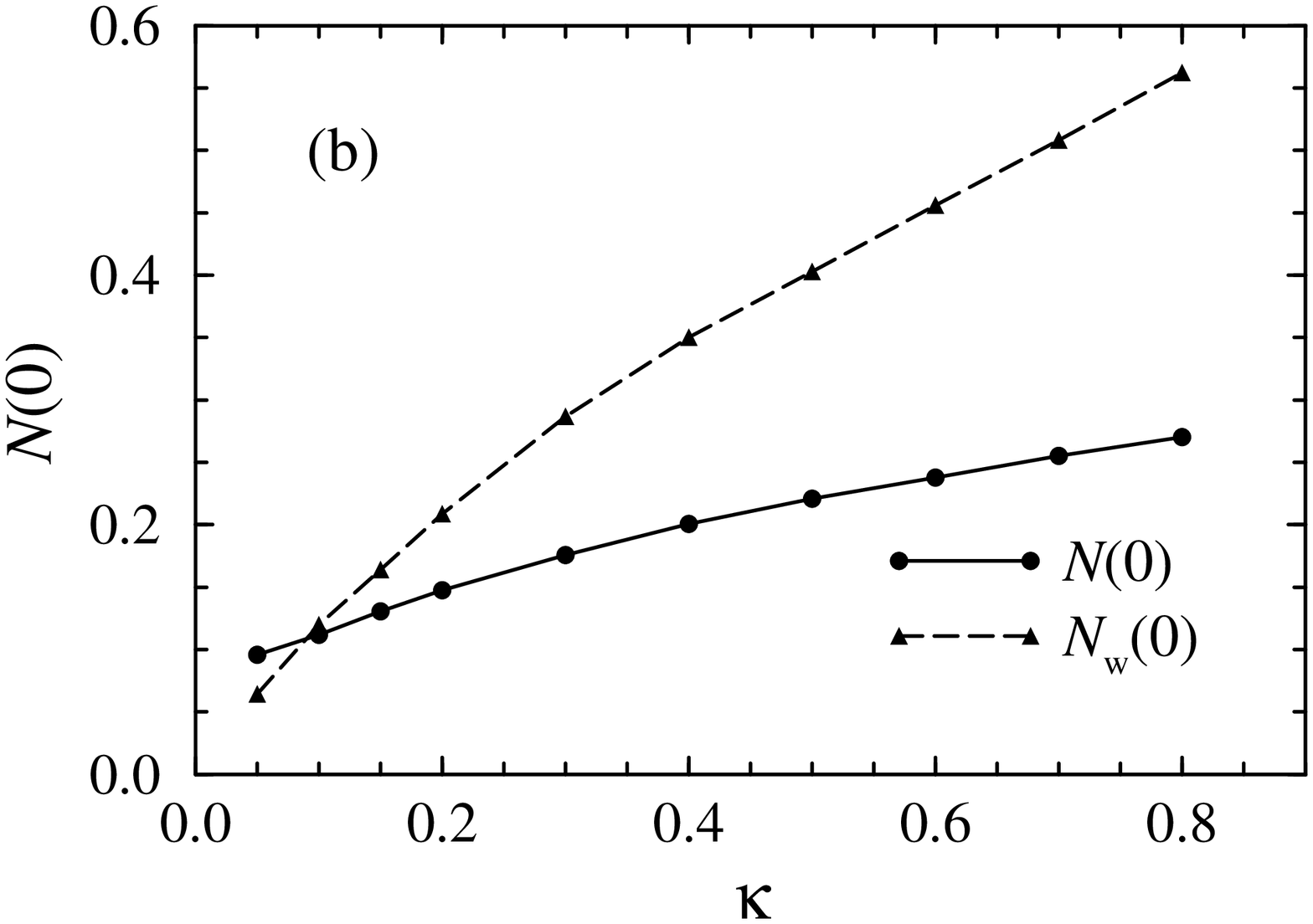,width=65mm,angle=-90,clip=}}
\caption{(a) Average QP weight $Z_{\rm av}$ and QP DOS ${\cal N}_{\rm
QP}$ vs. $\kappa$. (b) DOS ${\cal N}(0)$ and weighted DOS ${\cal
N}_{\rm w}(0)$ vs. $\kappa$.}
\end{figure}

\noindent
\begin{figure}[htb]
\center{\epsfig{file=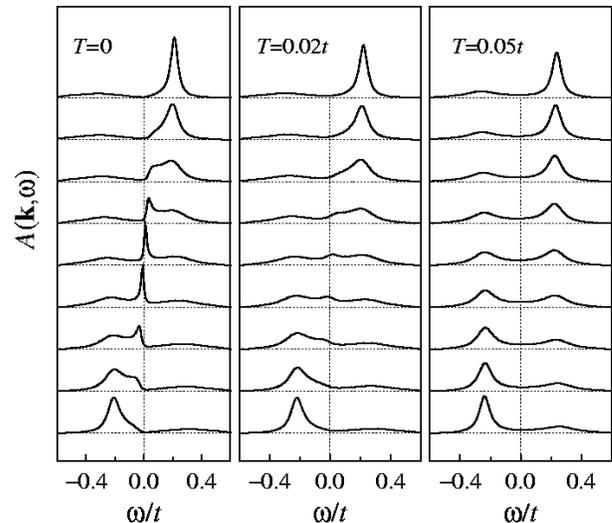,width=70mm,angle=-90,clip=}}\\
\caption{$A({\bf k},\omega)$ for $\kappa=0.1$ and ${\bf k}$ along the
central part of the line c in Fig.~2 for various $T$: $T=0$, $T=0.02
t$, and $T=0.05 t$. Momentum ${\bf k}$ ranges from $(\pi/4,3 \pi/4)
\mp (\pi/32,\pi/32)$.}
\end{figure}

\section{Self consistent calculation}

The full selfconsistent set of equations for $\Sigma=\Sigma_{\rm
pm}+\Sigma_{\rm lf}$, Eqs.(\ref{eq22}),(\ref{eq24}), and for $G$,
Eq.(\ref{eq15}), is solved numerically.  For given $\mu$ the FS
emerges as a solution determined by the relation $\zeta_{{\bf k}_F} +
\Sigma'({{\bf k}_F},0)= \mu$.  We should note that at given $\mu$,
electron concentration $c_e$ as calculated from the DOS
${\cal N}(\omega)$, Eq.(\ref{eq31}), does not in general
coincide with the one evaluated from the FS volume, $\tilde c_e=V_{\rm
FS}/V_0$.  Nevertheless, apart from the fact that within the $t$-$J$
model validity of the Luttinger theorem is anyhow under question
\cite{putt}, in the regimes of large FS both quantities appear to be
quite close. The position of the FS is mainly determined by
$\zeta_{\bf k}$ and $\Sigma_{\rm pm}$, while in this respect
$\Sigma_{\rm lf}$ is less crucial.

As discussed in Sec.~III we use in Eq.~(\ref{eq7}) the most
appropriate and simplest approximation to insert the unrenormalized
$A^0({\bf k},\omega)$, i.e., the spectral function without a
self-consistent consideration of $\Sigma_{\rm lf}$ but with
$\Sigma_{\rm pm}$ fully taken into account. We choose here $t'=-0.2t$
and again $\kappa= \sqrt{c_h}$ while $\eta_1$ and $\eta_2$ are
determined as a function of $c_h$ from model calculations
\cite{bonc}. We use $N=40\times40$ points in the Brillouin zone and
broadening $\epsilon/t=0.05$.

In Fig.~8 we present hole concentration $c_h$ vs. $\mu$ as obtained
from ${\cal N}(\omega)$ at $T=0$.  We solve selfconsistent equations by
iteration, whereby for $0.06 < c_h < 0.11$ we find in the equations an
instability signaled by oscillatory behavior instead of the convergence
and an unique solution can not be obtained in the region indicated by
the dashed line. However, at lower (and higher) doping the solution is
converged. It seems that the region of instability coincides with the
transition from the large to a small FS.

\noindent          
\begin{figure}[htb]
\center{\epsfig{file=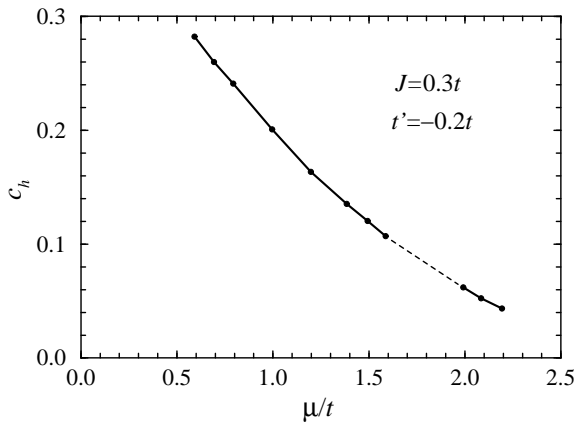,width=60mm,angle=-90,clip=}}
\caption{Hole doping $c_h$ as a function of the chemical potential
$\mu/t$, following from the selfconsistent calculation.}
\end{figure}   

The shape of the FS is most clearly presented with contour plot of the
electron momentum distribution function defined as
\begin{equation}
\tilde n({\bf k})=
\alpha^{-1}\int_{-\infty}^0
A({\bf k},\omega){\rm d} \omega.
\end{equation}
Results for a characteristic development of the FS with $c_h$ are
shown in Fig.~9. At higher doping, $c_h=0.26$ and also $c_h=0.20$, we
get a common large FS topology. In the intermediate doping regime,
$c_h=0.14$, the pseudogap is pronounced at momenta around $(\pi,0)$
and the FS shows a tendency of forming a small FS. The gap is more
pronounced because of longer AFM correlation length $\xi$ (smaller
$\kappa$). At $c_h <c_h^0 \sim 0.06$ solutions are consistent with a
small pocket-like FS whereby this behavior is enhanced by $t'<0$ as
realized in other model studies \cite{tohy}.  On increasing doping the
FS rather abruptly changes from a small into a large one as suggested
from the results of the SCBA \cite{ram00}.  The smallness of $c_h^0$
has the origin in quite weak dispersion dominated by $J$ and $t'$ at
$c_h \to 0$ which is overshadowed by much larger $\zeta_{\bf k}$ at
moderate doping, where the FS is large and its shape is controlled by
$t'/t$.

\noindent
\begin{figure}[htb]
\center{\epsfig{file=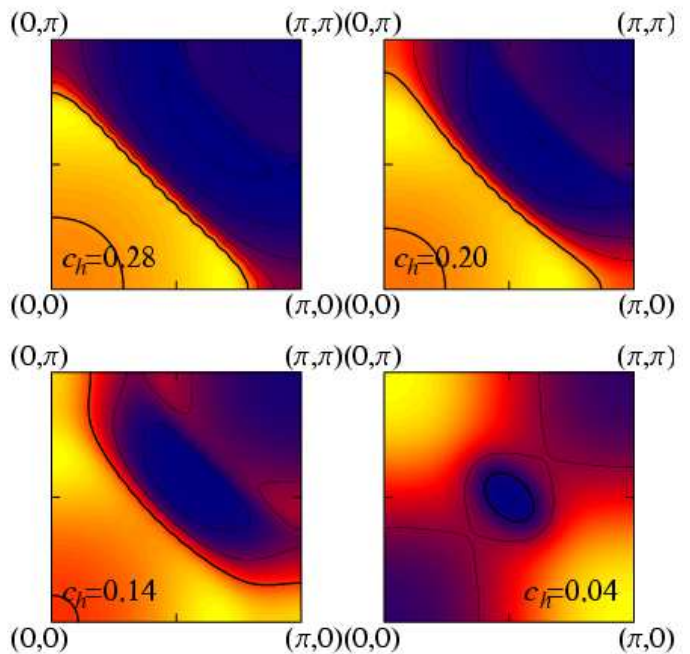,width=90mm,angle=-0,clip=}}\\
\caption{(color) Electron momentum distribution $\tilde n({\bf k})$
for various $c_h$. Thin contour lines represent $\tilde n({\bf k})$ in
increments $0.1$ while heavy line is corresponding to $\tilde n({\bf
k})=0.8$ for $c_h=0.26$, $0.20$, $0.14$ and $\tilde n({\bf k})=0.85$
for $c_h=0.04$.}
\end{figure}

In Fig.~10(a) and Fig.~10(b) we present calculated $A({\bf k},\omega)$
along the principal directions in the Brillouin zone, i.e.,
$(\pi/2,\pi/2) \to (0,0) \to (\pi,0)$.  It is evident that
$\Sigma_{\rm pm}$ leads to a strong damping of hole QP and quite
incoherent momentum-independent spectrum $A({\bf k},\omega)$ for
$\omega \ll -J$ which qualitatively reproduces ARPES and numerical
results \cite{jprev}.  Electron QP (at $\omega>0$) are in general very
different, i.e., with much weaker damping arising only from
$\Sigma_{\rm pm}$. Note a relatively high QP velocity in the higher
doping regime $c_h=0.26$, Fig.~10(a), as compared to a more narrow
dispersion on the scale $2J$ at low doping $c_h=0.04$ , Fig.~10(b),
where we find the regime of small pocket-like FS.
   
\vskip -0mm
\noindent   
\begin{figure}[htb]
\center{\epsfig{file=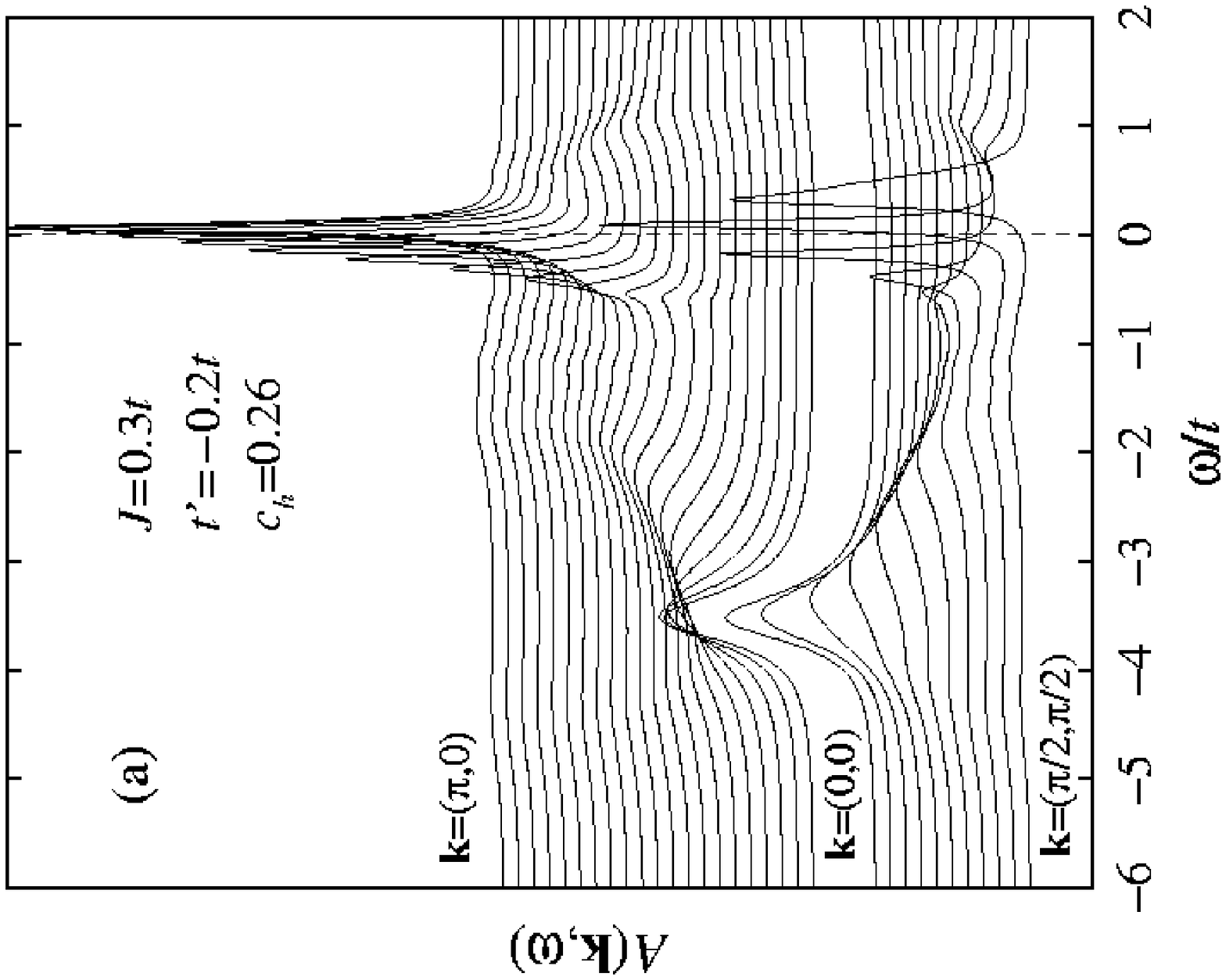,height=65mm,angle=-90,clip=}}\\[-0mm]
\vskip -0 mm
\center{\epsfig{file=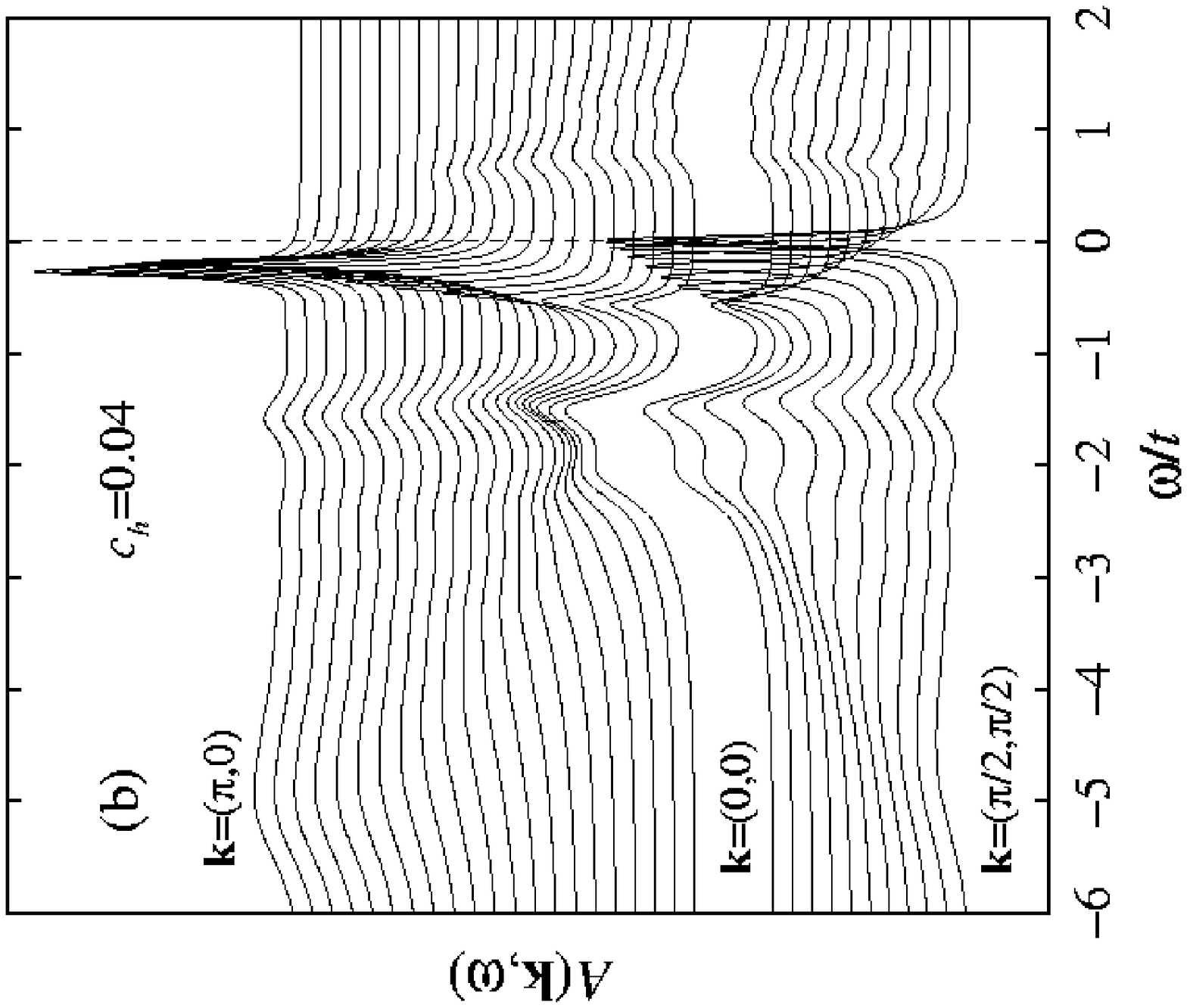,height=65mm,angle=-90,clip=}}\\
\caption{$A({\bf k},\omega)$ along main directions in the Brillouin
zone.}
\end{figure}

In Fig.~11 we present the development of spectral functions at fixed
$c_h=0.2$, but now varying $\kappa$ as an independent parameter.  Let
us concentrate on the emergence of the pseudogap near $(\pi,0)$.  At
$\kappa=0.4 \sim \sqrt{c_h}$ the pseudogap is essentially not yet
developed. Nevertheless, the gap opens with decreasing $\kappa$, in
particular for (at this doping unrealistic value) $\kappa=0.05$.

\noindent      
\begin{figure}[htb]   
\center{\epsfig{file=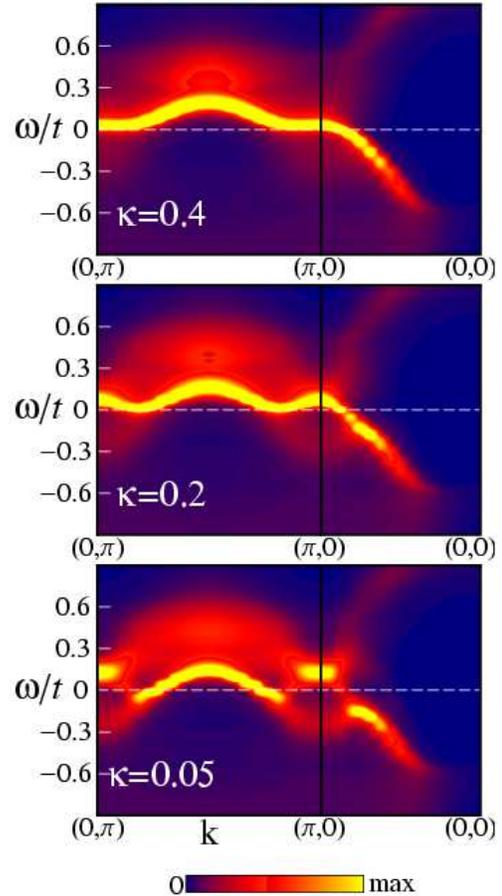,width=70mm,angle=-0,clip=}}\\
\caption{(color) Contour plot of $A({\bf k},\omega)$ for fixed doping
$c_h=0.2$ and various $\kappa$.}
\end{figure}

\section{Conclusions}

We have presented the theory of spectral functions within the $t$-$J$
model whereby our method is based on the EQM for projected fermionic
operators and on the decoupling approximation for the self energy
assuming the fermions and spin fluctuations as essential coupled
degrees of freedom. We first make some comments on the method:

\noindent a) The EQM approach for spectral functions (as well as for
other dynamical quantities) seem to be promising since it can treat
exactly the constraint which is essential for the physics of strongly
correlated electrons.

\noindent b) In finding the proper approximation for the self energy
within the EQM approach it is plausible that the main coupling is of
fermions with spin fluctuations, where close to the AFM
ordered state both transverse and longitudinal spin fluctuations are
important.  It is well possible, however, that other contributions
could be important, e.g., the coupling to pairing fluctuations.

\noindent c) In an ordered AFM our method reproduces naturally the
results for the spectral function (of a hole) within SCBA, which is
highly nontrivial, since both approaches are quite different.

\noindent d) The coupling to longitudinal spin fluctuations appear to
be most important for QP near the AFM zone boundary and is responsible
for the opening of the pseudogap. Here the coupling is only moderately
strong and can be treated in the lowest-order decoupling scheme.

\noindent e) The present theory uses the spin response as an input.
$\chi''({\bf q},\omega)$, Eq.(\ref{eq29}), corresponds in general to a
Fermi liquid or to a short-range AFM liquid. Results remain in fact
qualitatively similar as far as $\chi''({\bf q},\omega)$ is
nonsingular. In the opposite case, e.g., if we would use the MFL form
as an input, the Fermi surface would still be defined, but QP would
have vanishing weight $Z \to 0$ \cite{prel}.

Let us further discuss some main results of the presented theory:

\noindent a) The fermion-paramagnon coupling as manifested in $\Sigma_{\rm 
pm}$ remains effective and strong even at moderate doping. The full
calculation shows that the coupling leads to a large incoherent part in
the hole part ($\omega<0$) of the spectral function, as well as to
the renormalized band $\epsilon^{\rm ef}_{\bf k}$.
 
\noindent b) The main consequence of $\Sigma_{\rm lf}$ due to the
coupling to longitudinal spin fluctuations is the appearance of the
pseudogap at $\kappa<\kappa^*$. The pseudogap opens predominantly
along the AFM zone boundary and its extent is qualitatively given by
Eq.(\ref{eq37}), dependent on $J$ and $t'$ but not directly on
$t$. Evidently the pseudogap has similarity to d-wave-like dependence
along the FS, for $t'<0$ being largest near the $(\pi,0)$ point.

\noindent c) How strong is the pseudogap effect depends mainly on
$\kappa$. At small $\kappa \ll \kappa^*$ parts of the Fermi surface
near $(\pi/2,\pi/2)$ remain well pronounced (for $t'<0$) while the
Fermi surface within the pseudogap is suppressed, i.e., QP have a small
weight $Z_F \ll 1$, in particular near zone corners
$(\pi,0)$.

\noindent d) The simplified analysis yields large Fermi surface,
although a truncated one, except at very small $\kappa\ll \kappa^*$
where $\Sigma_{\rm lf}$ by itself induces a small hole-pocket-like
Fermi surface.  On the other hand, $\Sigma_{\rm pm}$ generates hole
pockets already for $c_h < c_h^0 \sim 0.06$. In fact, an instability
of the self consistent calculation indicates the emergence of
hole-pocket Fermi surface even at $c_h \alt 0.1$. However, it is well
possible that within the present approximation scheme $\Sigma_{\rm
pm}$ is overestimated at intermediate $c_h$, an indication for that
being quite a weak dispersion $\epsilon^{\rm ef}_{\bf k}$.
 
\noindent e) Our method is approximative in the evaluation of
$\Sigma$, hence it is not surprising that the volume of the Fermi
surface does in general not coincide with the one following the
Luttinger theorem. It is anyhow questionable if such a relation should
be valid within the $t$-$J$ model \cite{putt} due to the projected
basis and strong correlations. Nevertheless, in the regime of a large
Fermi surface the full calculation yields the Fermi surface volume
quite close to the Luttinger one.

\noindent f) For $\kappa<\kappa^*$ the QP within the pseudogap have
small weight $Z_F \ll 1$ but not diminished $v_F$, which is the
effect of the nonlocal character of $\Sigma({\bf k},\omega)$. A
consequence is that QP within the pseudogap contribute much less to QP
DOS ${\cal N}_{QP}$. This can explain the reduction of the latter with
doping and the appearance of the pseudogap in the specific heat being
essential for the understanding of the specific heat in underdoped
cuprates.

\noindent g) Although most results are presented for $T=0$, we can
discuss some effects of $T>0$. First effect is that within the
pseudogap the QP with $Z_F \ll 1$ are washed out (not just overdamped)
already for very small $T<T^s \ll \Delta^{PG}$. On the other hand, the
pseudogap is mainly affected by $\kappa$. So we can argue that the
pseudogap would be observable for $\kappa(c_h,T)<\kappa^* \sim
0.5$. This effectively determines the pseudogap crossover temperature
$T^*(c_h)$. From the quantitative studies of the $t$-$J$ model
\cite{soko,sing} it appears that in the region of interest $\kappa$ is
nearly linear in both $T$ and $c_h$ so we would get approximately
\begin{equation}
T^* \sim T^*_0 (1- c_h/c_h^*), \label{eq46}
\end{equation}
where $T^*_0 \sim 0.6 J$ and $c_h^* \sim 0.15$. 

Finally we make some comments on the relevance of our results to
experiments on cuprates, in particular with respect to observed
pseudogap and Fermi surface features:

\noindent a) The (large) pseudogap scale shows in ARPES on BSCCO as a
hump at $\sim 100$eV \cite{mars}. Our results indicate quite a similar
pseudogap scale, e.g. in Fig.~5 the $\omega<0$ pseudogap $\sim 0.3~t$
(note that $t \sim 0.4$eV), since $\Delta^{PG}$ is determined mainly
by $J$ and $t'$.

\noindent b) The truncated Fermi surface in underdoped BSCCO appears
as an arc (part of the large Fermi surface corresponding to $t'<0$) in
the Brillouin zone \cite{norm}, effectively not crossing the AFM zone
boundary, which is also characteristic of our results for
$\kappa<\kappa^*$, originating from the strong coupling to spin
fluctuations with commensurate $(\pi,\pi)$. The same is the origin of
shadow features in spectral functions pronounced at intermediate
doping and in particular at weak doping.

\noindent c) Our results for the depletion of the DOS ${\cal N}(0)$
(${\cal N}_w(0)$) with decreasing doping are qualitatively consistent
with the integrated PES (so far known for LSCO \cite{ino}) and STM
\cite{renn}, although in this relation the importance of matrix
element corrections is not yet clarified. The same holds for the
calculated decrease of QP DOS ${\cal N}_{QP}$ with doping essential in
connection with the specific-heat pseudogap in underdoped cuprates
\cite{lora}. It should be however mentioned that our results for both
${\cal N}(0)$ as well as of ${\cal N}_{QP} \propto \gamma$ indicate on
weaker suppression with decreasing doping than observed in
experiments. This is due to remaining contribution of Fermi surface
arcs, which could be overestimated in our approach for $\kappa \ll
\kappa^*$.

\noindent d) Both the value and the dependence of the pseudogap
temperature $T^*(c_h)$, as estimated in Eq.(\ref{eq46}), seem to be
very reasonable in connection with experimental evidence, arising from
various transport and magnetic properties in cuprates \cite{imad}.


\end{document}